\documentclass[preprint,showpacs,preprintnumbers,aps]{revtex4}
\usepackage{epsfig}
\usepackage{dcolumn} 
\usepackage{bm} 
\usepackage{latexsym}
\input {colordvi}
\def\mean#1{\left<#1\right>}
\def\Journal#1#2#3#4#5{{\it (#5) {#1} {#2} {#3}: #4 }}

\def\NPA{{Nucl. Phys. A}}

\def\PLB{{Phys. Lett. B}}

\def\PRL{Phys. Rev. Lett.\ }
\def\PRD{{Phys. Rev. D}}
\def\PRC{{Phys. Rev.  C}}

\begin{document}
\vspace*{-0.36in}
\title{How to Measure Specific Heat \\ Using Event-by-Event Average $p_T$ Fluctuations.} 
\author{ M.~J.~Tannenbaum}
\altaffiliation{Research supported by U.S. Department of 
Energy, DE-AC02-98CH10886.}
\affiliation{Brookhaven National Laboratory\\ Upton, NY 11973-5000 USA\\ for the PHENIX Collaboration\\ {\tt Email: mjt@bnl.gov}}
\date{\today}
\begin{abstract}
A simple way to visualize event-by-event average $p_T$ fluctuations is by assuming that each collision has a different temperature parameter (inverse $p_T$ slope) and that the ensemble of events has a temperature distribution about the mean, $\langle T\rangle$, with standard deviation $\sigma_T$. PHENIX characterizes the non-random fluctuation of $M_{p_T}$, the event-by-event average $p_T$,  by $F_{p_T}$, the fractional difference of the standard deviation of the data from that of a random sample obtained with mixed events. This can be related to the temperature fluctuation:
\[F_{p_T}=\sigma^{\rm data}_{M_{p_T}}/\sigma^{\rm random}_{M_{p_T}}-1\simeq(\langle  
n \rangle -1) \sigma^2_{T}/\langle T\rangle^2 \qquad .\] Combining this with the Gavai, {\it et al.},\cite{Gavai05} and Korus, {\it et al.},\cite{Korus} definitions of the specific heat per particle, a simple relationship is obtained:
\[ c_v/T^3={\mean{n}\over \mean{N_{tot}}} {1\over F_{p_T}} \qquad .\]
$F_{p_T}$ is measured with a fraction 
$\mean{n}/\mean{N_{tot}}$ of the total particles produced, a purely 
geometrical factor representing the fractional acceptance, $\sim 1/33$ in PHENIX. Gavai, {\it et al.} predict that $c_v/T^3=15$, which corresponds to 
$F_{p_T}\sim 0.20$\% in PHENIX, which may be accessible by measurements of $M_{p_T}$ in the range $0.2\leq p_T\leq 0.6$ GeV/c.
In order to test the Gavai, {\it et al.} prediction that $c_v/T^3$ is reduced in a QGP compared to the ideal gas value (15 compared to 21), precision measurements of $F_{p_T}$ in the range 0.20\% for $0.2\leq p_T\leq 0.6$ GeV/c may be practical. 
\end{abstract}
\maketitle

	\section{Introduction}
\begin{figure}[htb]
\begin{center}
\includegraphics[scale=1.2,angle=0]{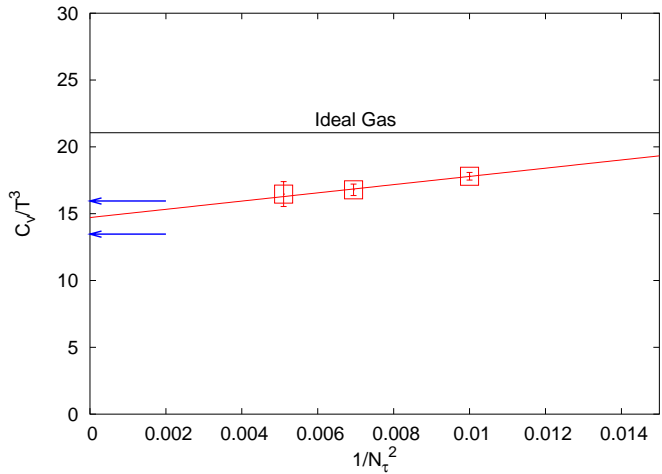}
\caption[]{Gavai, {\it et al.,} prediction for $c_v/T^3$~\cite{Gavai05}. \label{fig:Gavai}}
\end{center}\vspace*{-12pt}
\end{figure}

	R. Gavai, S. Gupta and S. Mukherjee~\cite{Gavai05} predict in ``quenched QCD" at $2T_c$ and $3T_c$ that the specific heat, $c_{V}/T^3$, differs significantly from the value for an ideal gas---15 compared to 21 (see Fig.~\ref{fig:Gavai}). Can this be measured?   
\section{Event-by-event average $p_T$ fluctuations and Specific Heat} 
\subsection{Single particle distributions}
\begin{figure}[htb]
\begin{center}
\includegraphics[scale=0.66,angle=0]{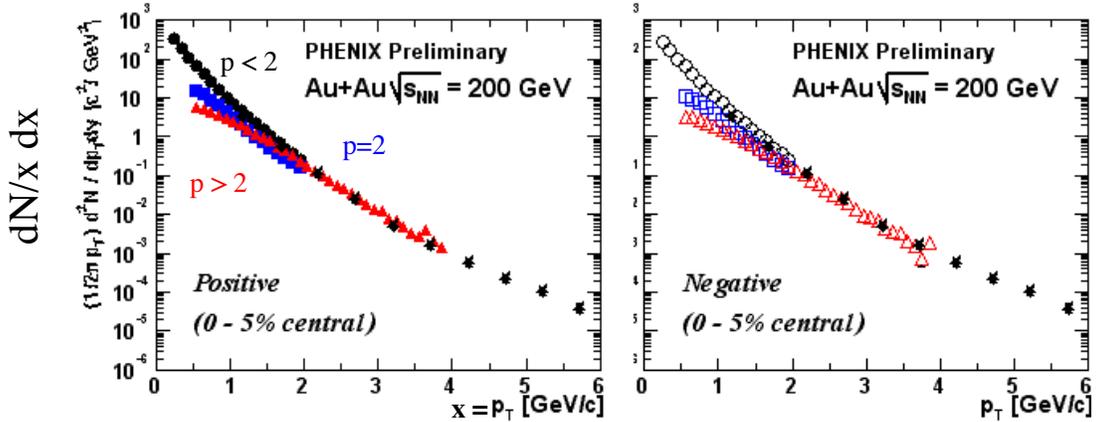}
\caption[]{Identified particle semi-inclusive invariant $p_T$ spectra for Au+Au central collisions~\cite{PX026}. At the lowest $p_T$, the $\pi^+$ are the highest, followed by $K^{+}$ and $p$ (left) and the same for the negatives (right). \label{fig:piKp}}
\end{center}\vspace*{-12pt}
\end{figure}

      The single particle transverse momentum ($p_T$) distribution averaged over all particles in all events for a p-p experiment (inclusive) or in all events of a given centrality class for an A+A experiment (semi-inclusive) is usually written in the form: 
\begin{equation}
{d\sigma\over {dp_T}}={b \over {\Gamma (p)}} {(b p_T)}^{p-1} e^{-b p_T} 
\qquad\mbox{or}\qquad
{d\sigma\over {p_T dp_T}}={{b^2} \over {\Gamma (p)}} {(b p_T)}^{p-2} e^{-b p_T} \qquad . 
\label{eq:ptdist}
\end{equation}
Equation~\ref{eq:ptdist} represents a Gamma 
distribution, where $\langle p_T\rangle=p/b$, $\sigma_{p_T}/\langle p_T\rangle = 1/\sqrt{p}$. Typically $b=6$ (GeV/c)$^{-1}$ and $p=2$ for p-p collisions. As shown in Fig.~\ref{fig:piKp}, the $p$ parameter depends on the particle type in central Au+Au collisions,  with $p< 2$ for $\pi^{\pm}$, $p\sim2$ for $K^{\pm}$ and $p >2$ for (anti-) protons, but the asymptotic slope tends to be the same for all particles. The `inverse slope 
parameter', $T=1/b$, is usually referred to as the `Temperature 
parameter'. 

\subsection{Event-by-Event Average}
 For events with $n$ detected charged particles with magnitudes of transverse momenta, $p_{T_i}$, the event-by-event average $p_T$, denoted $M_{p_T}$, is defined as: 
 \begin{equation}
M_{p_T}=\overline{p_T}={1\over n} \sum_{i=1}^n p_{T_i}\qquad .\label{eq:defMpT}
\end{equation}
By definition $\mean {M_{p_T}}\equiv \mean{p_T}=\mu$; however, it takes hard work to make one's data follow this identity to high precision ($\ll 1\%$). The standard deviation of $M_{p_T}$ is defined the usual way:
\begin{equation}
\sigma^2_{M_{p_T}}\equiv\mean{M^2_{p_T}}-\mean{M_{p_T}}^2 
={1 \over n^2}\left(n \sigma^2_{p_T}+\sum_{i=1}^n \sum_{j=1, j\neq i}^n \left\langle {(p_{T_i} -\langle p_T\rangle)  (p_{T_j}-\langle p_T\rangle)}\right\rangle\right) 
\qquad . 
\label{eq:defsigmaM}
\end{equation} 
If all the $p_{T_i}$ on all events are random samples of the same $p_T$ distribution, then: 
\begin{equation}
\sigma^2_{M_{p_T}}={\sigma^2_{p_T} \over n} \qquad ,
\label{eq:MpTrandom}
\end{equation}
where $\sigma_{p_T}=\sqrt{\mean{p_T^2}-\mean{p_T}^2}$ is the standard deviation of Eq.~\ref{eq:ptdist}, the inclusive $p_T$ spectrum (averaged over all events). 

    A nice illustration of what can be revealed by the event-by-event average that is not shown by the inclusive average over all events was given by Korus, {\it et al.}~\cite{Korus}. Suppose that each collision has a different temperature parameter such that the ensemble of events has a mean, $\mean{T}$, with standard deviation, $\sigma_{T}=\sqrt{\mean{T^2}-\mean{T}^2}$, about the mean. It is easy to show that for  this case: 
    \begin{equation}
    {{\sigma^2_{M_{p_T} }}\over \mu^2}-{1\over n}{\sigma^2_{p_T} \over \mu^2}=(1-{1 \over n}) {\sigma^2_T \over \mean{T}^2} \qquad. 
    \label{eq:MpT-sTT}
    \end{equation}
\subsection{Specific Heat}
   As pointed out by Korus, {\it et al.,}\cite{Korus} if the parameter $T$ would  correspond to the actual temperature of the system, not just the inverse slope of the $p_T$ distribution, then a basic equation of thermodynamics would relate the temperature fluctuations of a system to its total heat capacity~\cite{Stodolsky,Shuryak,Rajagopal}:
\begin{equation}
{1 \over C_V}={{\sigma^2_T }\over {\mean{T}^2}} \qquad ,    
\label{eq:CV1}
\end{equation}
where $C_V$ is an extensive quantity corresponding to the total number of particles in the system, $\mean{N_{tot}}$.  Thus the specific heat per particle is $c_V=C_V/\mean{N_{tot}}$. Gavai, {\it et al.,}\cite{Gavai05} refer to this same (dimensionless) quantity as $c_v/T^3$, resulting in the final equation:    
   \begin{equation}
   {c_{v}\over T^3}={1\over {\langle N_{tot}\rangle}}\frac{1}{\sigma^2_T/\langle T\rangle^2}
   \label{eq:cvT3sTT}
   \end{equation}
   where $n$ represents the number of particles used in the calculation of $M_{p_T}$ (Eq.~\ref{eq:MpT-sTT}) from which $\sigma_T/\mean{T}$ is determined. 
\section{Measurements of $M_{p_T}$}
\begin{figure}[hbt]
\begin{center}
\begin{tabular}{cc}
\includegraphics[scale=0.9,angle=0]{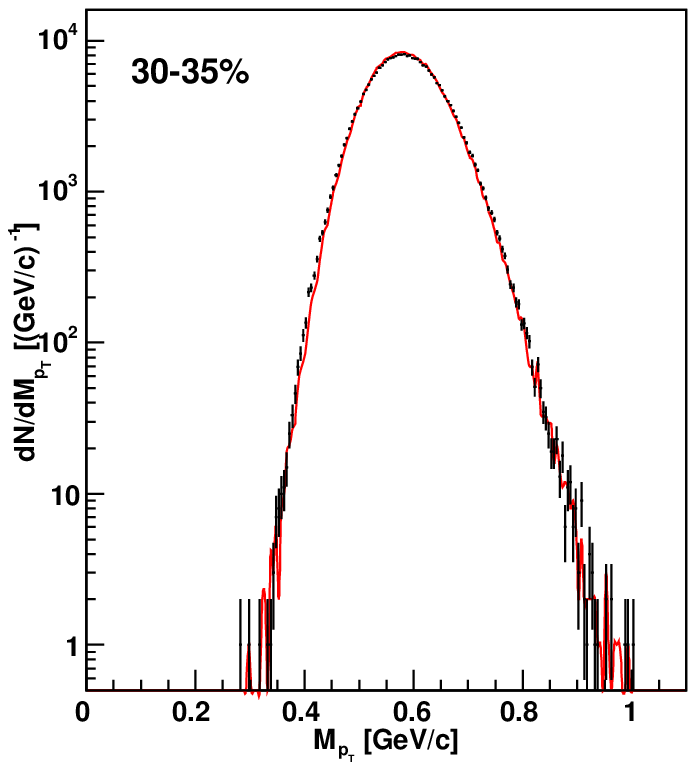}&
\includegraphics[scale=0.9,angle=0]{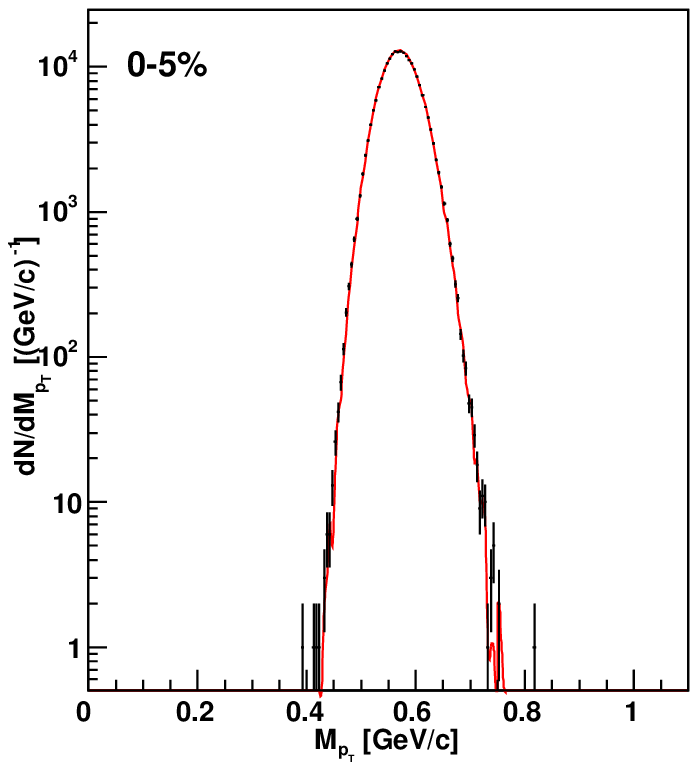}
\end{tabular}
\end{center}\vspace*{-0.25in}
\caption[]{$M_{p_T}$ for 30-35\% and 0-5\% centrality classes\cite{PX200}: data (points) mixed-events (histogram). \label{fig:MpT}}
\end{figure}

        The measured $M_{p_T}$ distributions for two centrality classes in  $\sqrt{s_{NN}}=200$ GeV Au+Au collisions in PHENIX~\cite{PX200} are shown in Figure~\ref{fig:MpT} (data points) compared to a random baseline (histograms).   Mixed-events are used to define the baseline for random fluctuations of $M_{p_T}$. This has the advantage of effectively removing any residual detector-dependent effects. The event-by-event average distributions
are very sensitive to the number of tracks in the event (denoted $n$), so the mixed event sample is produced with the {\em identical} $n$ distribution as the data. Additionally, no two tracks from the same data event are
placed in the same mixed event in order to remove any intra-event correlations in $p_T$. Finally, $\langle M_{p_T}\rangle$ must exactly match the semi-inclusive $\langle p_T\rangle$.

\begin{figure}[hbt]
\begin{center}
\begin{tabular}{cc}
\includegraphics[scale=0.4,angle=0]{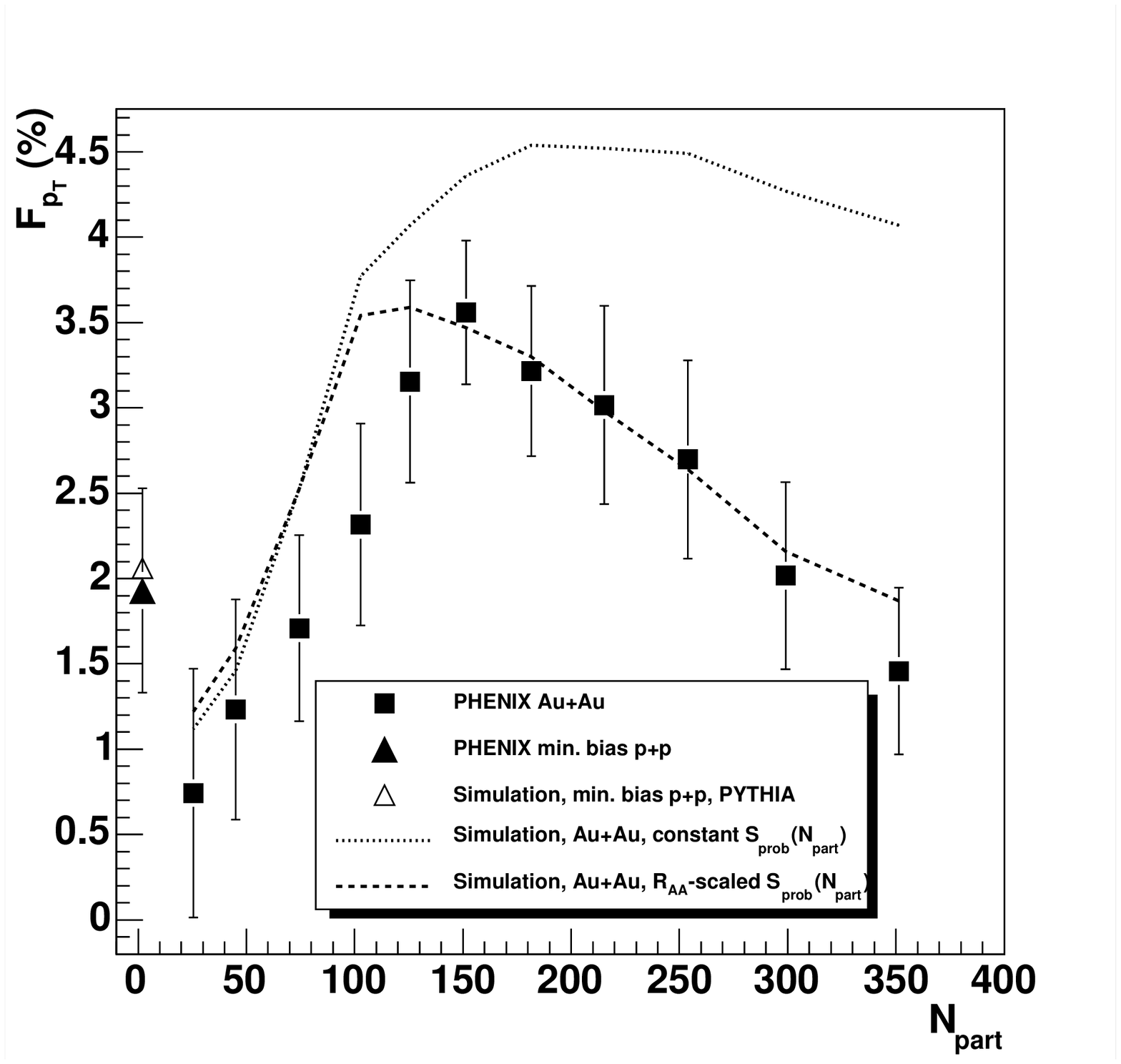}&
\includegraphics[scale=0.4,angle=0]{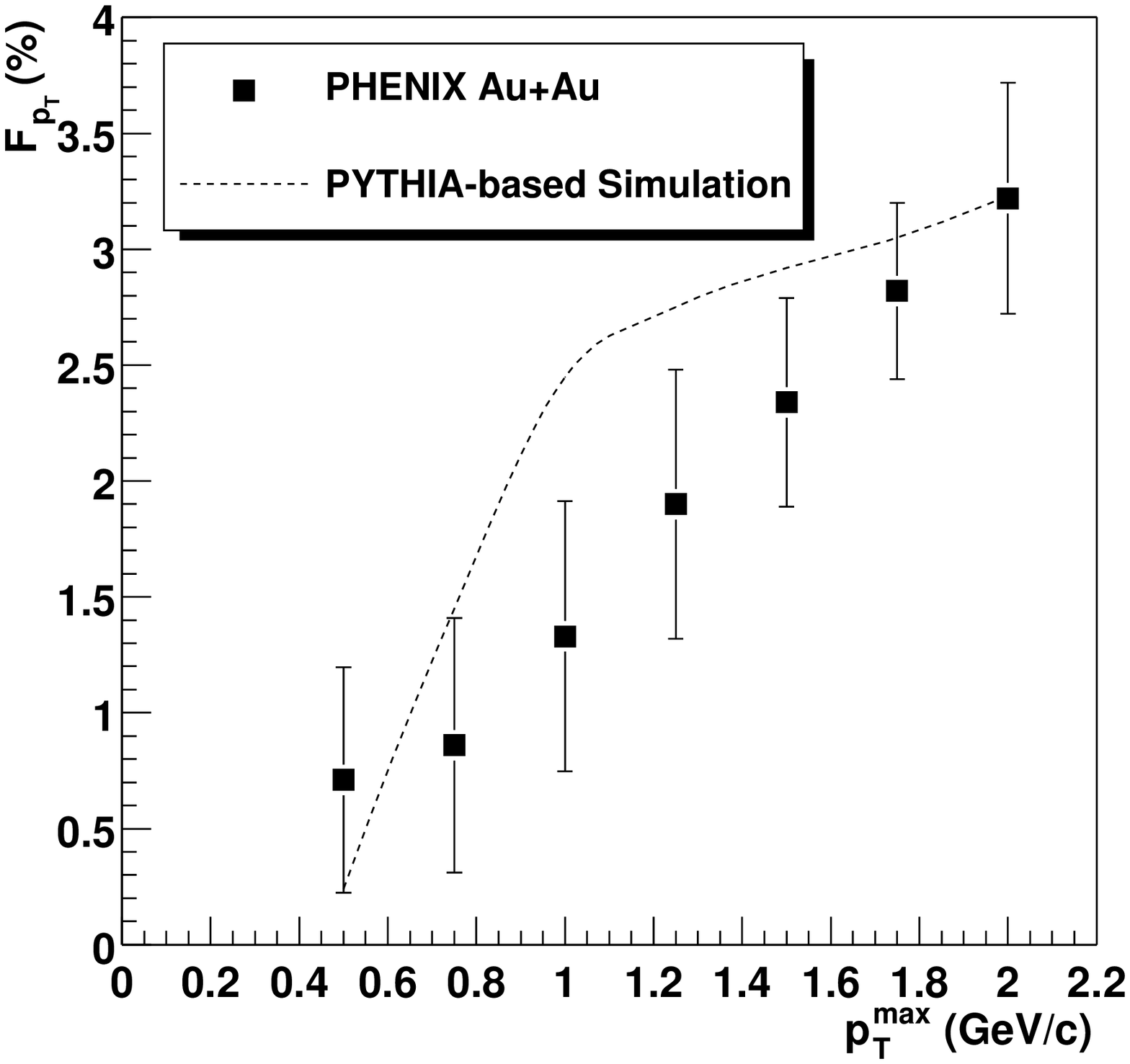}
\end{tabular}
\end{center}\vspace*{-0.25in}
\caption[]{$F_{p_T}$ vs centrality and $p_{T}^{\rm max}$ compared to simulations~\cite{PX200}.  \label{fig:Ft}}
\end{figure}

	The non-Gaussian, Gamma distribution shape of the $M_{p_T}$ distributions is evident.
The difference between the data and the mixed-event random baseline distributions is barely visible to the naked eye. PHENIX quantifies the non-random fluctuation by the fractional  difference of the standard deviations of $M_{p_T}$ for the data and the mixed-event (random) sample:
	\begin{equation}
	F_{p_T}\equiv \frac{\sigma_{M_{p_T},{\rm data}}-\sigma_{M_{p_T},{\rm mixed}}}{\sigma_{M_{p_T},{\rm mixed}}} \qquad ,
	\label{eq:def:FpT}
	\end{equation}
	which is on the order of a few percent. 
The results are shown (Fig.~\ref{fig:Ft}-left) as a function of centrality (represented by $N_{part}$) for charged particle tracks in the range \mbox{0.2 GeV/c $\leq p_T\leq 2.0$ GeV/c}; and, for the 20-25\% centrality class ($N_{part}=181.6$), over a varying $p_T$ range, \mbox{0.2 GeV/c $\leq p_T\leq p_T^{\rm max}$} (Figure~\ref{fig:Ft}-right). 
The steep increase in $F_{p_T}$ for the small increase in the number of tracks with increasing $p_{T}^{\rm max} > 1$ GeV/c is consistent with correlations due to jet production as shown by the dotted lines~\cite{PX200}. However, other explanations have been proposed~\cite{Ferreiro}. Note that the errors are entirely systematic, due to time-dependent detector variations. Comparatively, statistical errors are negligible. 
\section{How to measure $c_v/T^3$.}
   For the small values of $F_{p_T}$ observed, one can make use of the identity 
   \begin{equation}
   {{\Delta\sigma^2} \over {\sigma^2}}=2 {{\Delta\sigma} \over {\sigma}}=2F
 \label{eq:identity}
 \end{equation}
 to obtain the relation:
 \begin{equation}
 (1-{1 \over n}) {\sigma^2_T \over T^2}
={{\sigma^2_{M_{p_T} }}\over \mu^2}-{1\over n}{\sigma^2_{p_T} \over \mu^2}=2F_{p_T} {1\over n}{{\sigma^2_{p_T}} \over \mu^2}={{2F_{p_T}}\over {np}}\simeq {F_{p_T} \over n}\qquad .
\label{eq:relate}
\end{equation}
Combining Eq.~\ref{eq:relate} with Eq.~\ref{eq:cvT3sTT}, we obtain the simple and elegant  expression:
\begin{equation}
{c_{v}\over T^3}={{\langle n\rangle}\over {\langle N_{tot}\rangle}}\frac{1}{F_{p_T}}
\qquad .
\label{eq:cvF}
\end{equation}
Note that $F_{p_T}$ is measured with a fraction $\mean{n}/\mean{N_{tot}}$ of the total particles produced, which is a purely geometrical factor representing the fractional acceptance of the measurement. For example, if all particles are produced in a range $\delta\eta_{\rm FWHM}$ (assuming a flat or trapezoidal $dn/d\eta$ over this interval) and if including the neutrals gives a factor of 1.5 more total particles than charged particles; and if $F_{p_T}$ is measured with charged particles in an acceptance $\delta\eta_c$, $\delta\phi_c/2\pi$, which due to the $p_T$ cut only represents a fraction $f_c$ of the charged particles on that solid angle, then:
\begin{equation}
 {\mean{N_{tot}} \over \mean{n}}=\frac{1.5\times 2\pi\times \delta\eta_{\rm FWHM}} {f_c \times \delta\phi_c \times \delta\eta_c } \qquad .
 \label{eq:fracN}
\end{equation}
    
\begin{figure}[htb]
 \begin{center}
 \includegraphics[scale=0.8,angle=0]{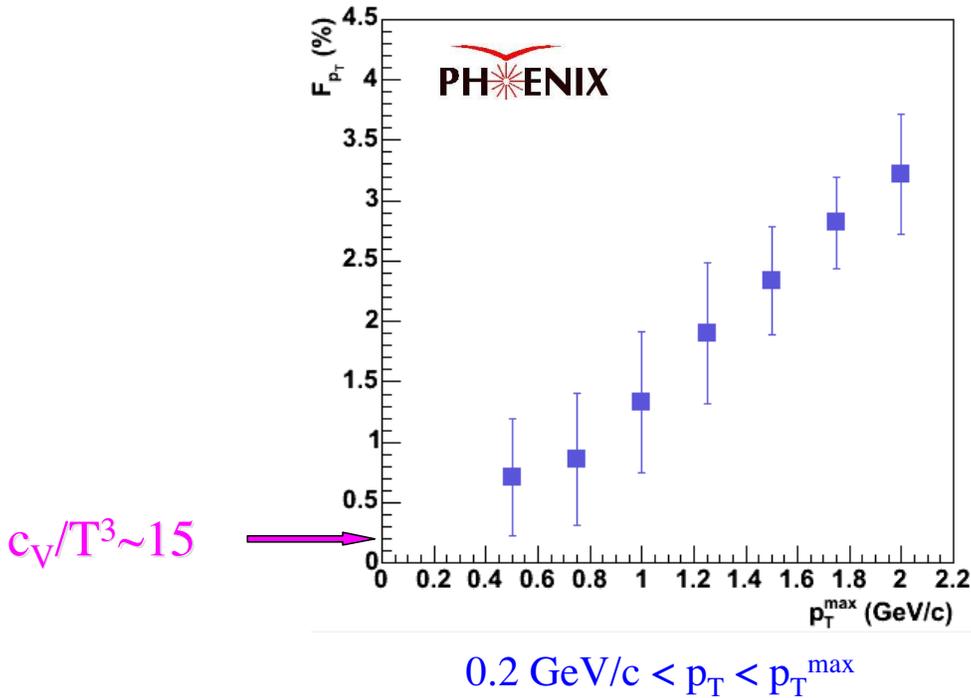}
\caption[]{Gavai {\it et al.}, prediction~\cite{Gavai05} compared to PHENIX measurement. \label{fig:Ft2}}
\end{center}\vspace*{-12pt}
\end{figure}

    For RHIC at $\sqrt{s_{NN}}=200$ GeV, $\delta\eta_{\rm FWHM}=\pm3.5$~\cite{PhobosWP}, and the PHENIX acceptance was $\delta\phi=\pi$, $\delta\eta=\pm0.35$, $f_c$=0.9 for $p_T\geq 0.2$ GeV/c, resulting in ${\mean{N_{tot}} / \mean{n}}=33$. From Fig.~\ref{fig:Ft}, $F_{p_T}$ is of order 2\% but most of that is due to jets, so the effect due to temperature fluctuations is $< F_{p_T}$, say 1\%,  
 so we obtain $c_v/T^3 > 1/(33 * 0.01)=3$. This is to be compared to the Korus, {\it et al.}~\cite{Korus}, result of $c_v/T^3=60\pm 100$~\cite{MJTcalc} from the NA49 data. Recall that Gavai, {\it et al.},~\cite{Gavai05} predict a value of 15 for $c_v/T^3$, which would correspond to a value of $F_{p_T}=1\% \times 3/15=0.20\% $ for the data in Fig.~\ref{fig:Ft} (see Fig.~\ref{fig:Ft2}). 
Perhaps this precision can be achieved by concentrating on the region $p_{T}^{\rm max}\leq 0.6$ GeV/c, where jets have least effect. Also, as the present error is totally systematic due to run-by-run variation, there is hope that a substantial reduction should be possible. 
\section{Conclusions} 
   In central heavy ion collisions, the huge correlations in p-p collisions are washed out~\cite{Shuryak}. The remaining correlations are: Jets; Bose-Einstein correlations; Hydrodynamic Flow. These correlations seem to saturate the present fluctuation measurements. No other sources of non-random fluctuations have been observed. This puts a severe constraint on the critical fluctuations that were expected for a sharp phase transition but is consistent with the present expectation from lattice QCD that the transition is a smooth crossover. In order to see the temperature fluctuations predicted by $c_v/T^3\simeq 15$ in lattice gauge calculations, present sensitivity needs to be improved by an order of magnitude by removing the known sources of correlation and improving the measurement errors. An interesting check of whether temperature fluctuations, rather than the correlations noted above, produce the observed non-random fluctuations is provided by Eq.~\ref{eq:relate}: for a pure $\sigma^2_T/\mean{T}^2$ fluctuation, $F_{p_T}$ for a given centrality should increase linearly with the number of tracks measured (e.g. by increasing the solid angle---PHENIX {\it cf.} STAR).


\begin{thebibliography}{99}
\bibitem{MJTcalc} Using the variable $\phi=\sigma_{p_T} F_{p_T}=\sqrt{2}\mean{T} F_{p_T}=0.6\pm 1.0 $ MeV for $T=180$, from NA49, cited by Korus, {\it et al.},~\cite{Korus}  we find the value  $F_{p_T}=0.24\%\pm 0.39\%$. The values for $\mean{n}=270$ and ${\mean{N_{tot}}/\mean{n}}=1.5\times 5$ also cited yield $c_v/T^3=1/(7.5 F_{p_T})=56\pm 93$ in agreement with the quoted value of $66\pm 100$.
\end{thebibliography}
\end{document}